\documentclass[aps,pra,floats,twocolumn,superscriptaddress,showpacs,showkeys,floatfix]{revtex4-1}
\usepackage[utf8]{inputenc}
\usepackage{comment}
\usepackage{epsfig}
\usepackage{xcolor}
\usepackage{units}
\usepackage{hyperref}
\usepackage{amsmath}
\usepackage{amsfonts}
\usepackage{amssymb}
\usepackage{physics}[abs]
 
\begin{document}

\title{Exploring phonon-like interactions in one-dimensional Bose-Fermi mixtures}
\author{Axel Gagge}
\affiliation{Department of Physics, Stockholm University, SE-106 91 Stockholm, Sweden}
\author{Th. K. Mavrogordatos}
  \affiliation{Department of Physics, Stockholm University, SE-106 91
  Stockholm, Sweden}
  \affiliation{ICFO -- Institut de Ci\`{e}ncies Fot\`{o}niques, The Barcelona Institute of Science and Technology, 08860 Castelldefels (Barcelona), Spain}
\author{Jonas Larson}
\email[e-mail address: ]{Jonas.Larson@fysik.su.se}
\affiliation{Department of Physics, Stockholm University, SE-106 91 Stockholm, Sweden}

\date{\today}

\begin{abstract}
With the objective of simulating the physical behavior of electrons in a dynamic background, we investigate a cold atomic Bose-Fermi mixture confined in an optical lattice potential solely affecting the bosons. The bosons, residing in the deep superfluid regime, inherit the periodicity of the optical lattice, subsequently serving as a dynamic potential for the polarized fermions. Owing to the atom-phonon interaction between the fermions and the condensate, the coupled system exhibits a Berezinskii-Kosterlitz-Thouless transition from a Luttinger liquid to a Peierls phase. However, under sufficiently strong Bose-Fermi interaction, the Peierls phase loses stability, leading to either a collapsed or a separated phase. We find that the primary function of the optical lattice is to stabilize the Peierls phase. Furthermore, the presence of a confining harmonic trap induces a diverse physical behavior, surpassing what is observed for either bosons or fermions individually trapped. Notably, under attractive Bose-Fermi interaction, the insulating phase may adopt a fermionic wedding-cake-like configuration, reflecting the dynamic nature of the underlying lattice potential. Conversely, for repulsive interaction, the trap destabilizes the Peierls phase, causing the two species to separate.
\end{abstract}

\pacs{05.30.Rt, 42.50.Ct, 75.10.Kt}
\keywords{Bose-Fermi mixture, Peierls instability, harmonic trap, lattice potential}
\maketitle

\section{Introduction}

Over the past decades, we have witnessed a rampant growth of experimental methods devised to cool and control dilute gases. The attainment of Bose-Einstein condensates (BECs)~\cite{anderson_observation_1995-1} was soon to be followed by in-depth explorations of BEC dynamics in light-induced periodic potentials~\cite{morsch_bloch_2001,denschlag_bose-einstein_2002}, paving the way for the groundbreaking demonstration of the Mott-superfluid phase transition. This exemplary quantum phase transition was predicted by the Bose-Hubbard model~\cite{fisher1989boson,jaksch1998cold} for an atomic gas loaded into an optical lattice~\cite{greiner2002quantum}. Since then, trapped and cooled dilute atomic gases have developed into a versatile laboratory where quantum matter can be studied in a controlled and detailed fashion, constituting one of the most promising platforms for the realization of analog quantum simulators~\cite{lewenstein2007ultracold,Bloch2008,daley2022practical}. 

\begin{figure}
    \centering
    \includegraphics[width=0.45\textwidth]{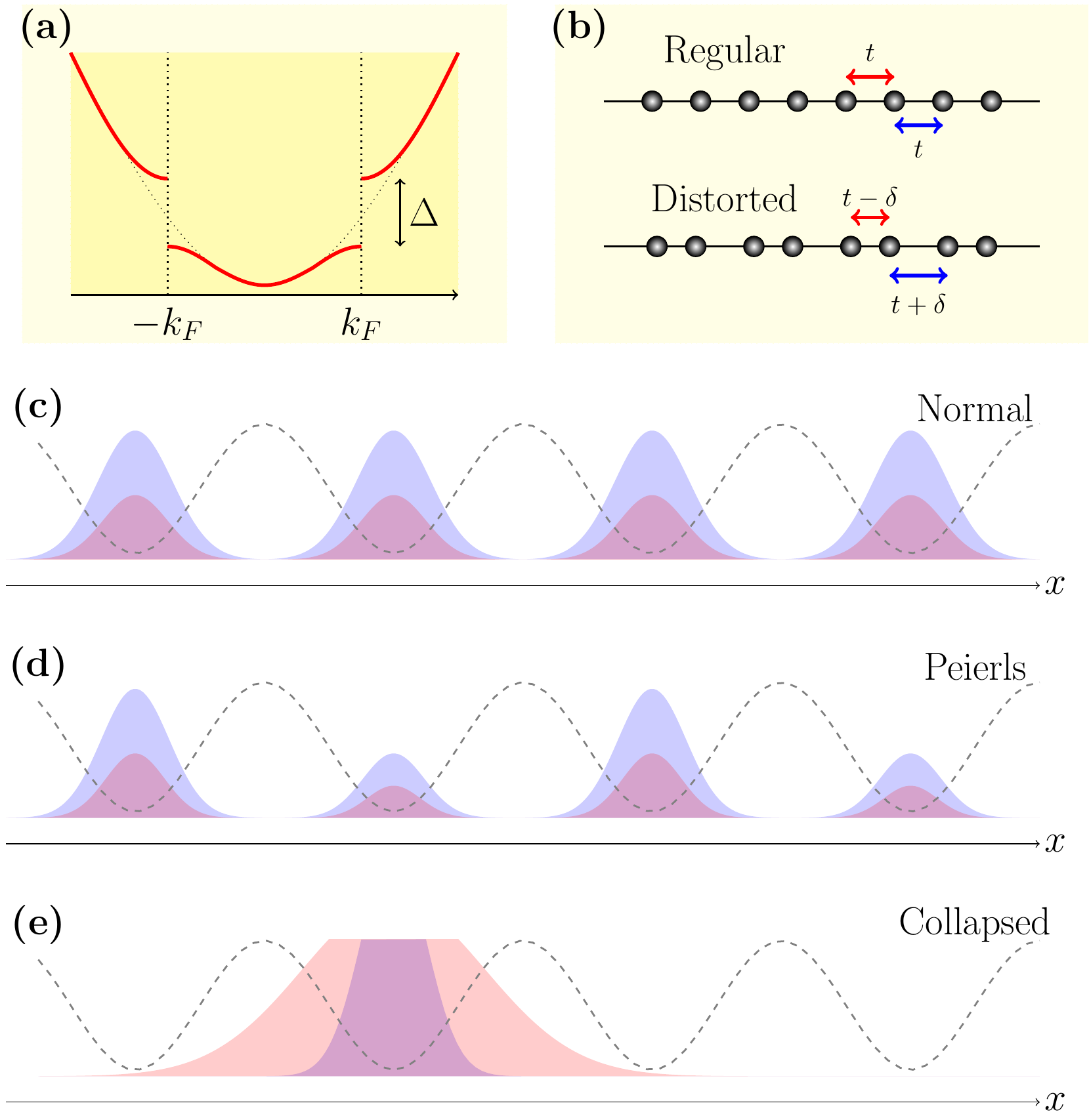}
    \caption{(color online) {\it Schematic description of the Peierls distortion.} At a fermion filling corresponding to a wavenumber $k_F$, opening up a gap of the dispersion at $k_F$ will lower the kinetic energy of the fermions, as depicted in {\bf (a)}. The gap opening results from a lattice distortion of the dynamical lattice, as shown in {\bf (b)}. Here, the period has been doubled through shifting every second atom by a distance $\delta$, and consequently, the size of the Brillouin zone has been halved. Thus, for half filling, we encounter a realization of the Su-Schrieffer–Heeger (SSH) model, while for other fillings, the lattice distortion will generate another periodicity. In the normal phase, the condensate will share the same periodicity as the lattice and predominantly populate the lattice minima {\bf (c)}. In the Peierls phase, for half filling, the bosonic density will alternate between every second site {\bf (d)}. Upon comparison of {\bf (b)} and {\bf (d)}, we note that the periodicity has been broken in two different ways: in {\bf (b)}, the densities are held fixed, but the locations of the sites have been shifted, while in {\bf (d)}, the locations are fixed and the densities have been altered. The former can be seen as a ‘phase modulation,’ and the latter as an ‘amplitude modulation’ of the densities. Finally, in {\bf (e)}, we depict the collapsed phase where all atoms, bosons, and fermions have been compressed to a small region of the lattice.}
    \label{fig:system}
\end{figure}

Many of the open questions in the field of quantum phase transitions, especially those revolving around the formation and characterization of exotic quantum phases of matter, cannot, however, be directly addressed by the Bose-Hubbard model {\it per se}. Although the classical laser field forming the optical lattice is, in principle, dynamic, the back-action between the trapped particles (modeled as beam splitters) and the lattice is typically very weak~\cite{asboth2007collective,asboth2008optomechanical}. One may then, to a very good approximation, treat the lattice as the outcome of applying a periodic static potential. Certainly, this approximation would be sufficient to emulate many paradigmatic lattice models. However, such classical potentials do not take into account any back-action between the lattice and the conducting matter, which we know is essential to explain several phenomena like the Peierls distortion~\cite{peierls_quantum_1955} [see Figs.~\ref{fig:system} (a) and (b)] and superconductivity~\cite{mahan2013many}. In solids, effects of that kind result from electron-phonon interactions emerging from the very nature of a dynamical lattice.

A central theme in our analysis is that assigning tractable degrees of freedom to the lattice renders its description dynamic and enables the simulation of some analog of phonon-like interactions. One possibility is to couple the atoms to the light field of an optical resonator~\cite{deng2014bose,pan2015topological}, where substantial Stark shifts are observed even in the presence of a few photons. However, in current experiments, only a few modes of the resonator actively participate in the light-matter interaction, and it is, therefore, not possible to locally modify the dynamical lattice. Instead, one must turn to multi-mode cavities~\cite{ballantine2017meissner}, which still pose difficulties in reaching configurations similar to those encountered in real solids~\cite{lewenstein2006travelling}. A viable alternative arises when considering atoms directly interacting with an ionic crystal~\cite{bissbort_emulating_2013-1}. Here, harmonically trapped ions form a Wigner crystal, while additional neutral atoms move within this lattice. This ion-atom system bears obvious similarities to a real solid, albeit being experimentally challenging. In fact, the crystalline structure is not necessary for exploring phonon-like interactions. In mixtures of different atomic species — either different atoms or different internal states of the same atomic species — the interplay between subsystems can lead to intriguing effects.

Coming now to single atomic species, the use of Feshbach resonances allows experimenters to control the strength, and even the sign, of all involved interactions~\cite{stan2004observation,best2009role,kawaguchi2012spinor,park2012quantum,ferrier-barbut_mixture_2014-1,desalvo_observation_2019}, aiming to probe the resulting phase diagram. In mixtures of bosons and spin-polarized fermions (hereinafter referred to as BF mixtures), it is well known that an attractive BF interaction leads to a so-called pairing phase in the strongly correlated regime. This phase has been studied for weak BF interactions, $g_{bf}$, in one dimension (1D)~\cite{cazalilla_instabilities_2003-1,miyakawa_peierls_2004-1,rizzi_pairing_2008-1}, as well as in two\cite{buchler_supersolid_2003-1,klironomos_pairing_2007-1} and three dimensions~\cite{titvinidze_supersolid_2008-1}. The phase in question collapses if the interaction becomes too strong, resulting in clumping of the atoms and breaking of translational invariance. The effect of optical lattices on BF mixtures has also been investigated in Refs.~\cite{albus_mixtures_2003-1,mathey_luttinger_2004-2,roth2004quantum,salerno2005matter,pazy_holstein_2005-1,bruderer_polaron_2007-1,lan_optical_2014-1}. For deep lattices and/or very strong interaction, such systems can be described by a BF Hubbard model as they enter an insulating phase of composite fermions~\cite{lewenstein_atomic_2004-1,pollet_phase_2006-1}. The physical behavior is typically described within a Wannier-basis expansion for both species, where the bosons can be construed as agents of effective onsite energy shifts for the lighter fermions. The imposed approximations in such a scheme omit certain back-action between the two subsystems in comparison to the self-consistent analysis we employ here. Nevertheless, it is still possible to encounter Peierls phases, supersolids, and charge density waves~\cite{buchler_supersolid_2003-1,lewenstein_atomic_2004-1}. Furthermore, if the repulsive boson-boson interaction is weak, the system can enter a regime of phase separation where the bosons and fermions completely avoid each other~\cite{buchler2004phase}.

Reporting from the experimental front, an early study focused on how the coherence of an atomic condensate –- held in place by a cubic optical lattice –- is affected by the presence of fermionic atoms~\cite{gunter2006bose}. Quantum degeneracy for both bosons and fermions was attained in Ref.~\cite{ferrier-barbut_mixture_2014-1}, where the two species were treated on equal footing, a trend followed in a series of subsequent papers~\cite{delehaye2015critical,roy2017two,desalvo2017observation,yao2016observation}. Further experimental investigations have also considered a particular regime emerging for a condensate in weak contact with much lighter fermionic atoms~\cite{desalvo_observation_2019,edri_observation_2020,desalvo2017observation}. Here, the fermions effectively induce a so-called Ruderman–Kittel–Kasuya–Yosida (RKKY) long-range boson-boson interaction~\cite{ruderman1954indirect}. For attractive Bose-Fermi interaction, it was demonstrated that a self-sustained trap may emerge for those fermions located inside the bosonic condensate~\cite{desalvo2017observation}, an effect which may as well lead to the formation of soliton trains~\cite{desalvo_observation_2019}. The fermion-mediated spin-spin interaction in a spinor condensate has also been recently observed by means of microwave spectroscopy~\cite{edri_observation_2020}.

Carrying on with the thread of recent investigations on collective phenomena brought about by a {\it coherent} interaction between Bose-Fermi atomic clouds, in this report, we study a 1D atomic BF mixture where only the bosons are subject to an optical lattice. The motivation for considering such a configuration is drawn from the resemblance to an actual solid – the optical lattice orders the bosons in a crystalline structure. Our spin-polarized fermions experience only the periodic potential arising from the boson atomic density but no externally imposed static potential, with the occasional exception of a harmonic trap. A similar idea, although experimentally more challenging, was proposed in Ref.~\cite{johnson2016hubbard}, where the crystalline order for the bosons was established from a rapidly rotating condensate that created a triangular Abrikosov vortex lattice. We focus on the limit of a weak boson-boson (BB) interaction and an optical lattice of amplitude $\leq 20 E_R$ ($E_R$ is the recoil energy), where the bosonic gas is expected to form a superfluid, see Fig.~\ref{fig:system}(c). In this regime, we can work within a mean-field approximation for the condensed bosons, which act as a classical dynamical lattice felt by the fermions. The coupled system is solved self-consistently such that every back-action, at the mean-field level, is taken into account (without imposing, for instance, any single-band nor tight-binding approximations). At this hybrid mean-field level, the mixture displays a rich phase diagram. For strong BF interaction, with a coupling rate $g_{bf}$, the system either experiences a collapse ($g_{bf}\gg0$), in which the two species overlap and populate only a small fraction of the lattice, or a separation ($g_{bf}\gg0$) where instead the two species avoid each other and populate different parts of the lattice. These are first-order transitions, even though translational invariance is spontaneously broken in both cases. For a non-zero interaction beyond a critical coupling $g_{bf}=g_{bf}^c$, an instability occurs such that the period of the state is different from that of the underlying lattice; for example, for half-filling, a periodic doubling is found. The system then transitions from a Luttinger liquid (LL) phase into a Peierls phase via a Berezinskii-Kosterlitz-Thouless (BKT) transition. The Peierls phase manifests itself through a non-zero gap $\Delta_P$ at the Fermi wavenumber $k_\mathrm{F}$, {\it i.e.} $\Delta_P \neq 0$ for $g_{bf}$ beyond the critical interaction strength $g_{bf}^c$.

We further present an account of the effects arising due to the presence of a confining harmonic trap. For a sufficiently strong repulsive boson-fermion interaction, it has already been demonstrated that, at the mean-field level, the trap induces a separation of the two species~\cite{nygaard1999component}. This instance suggests ruling out a Peierls-like phase when a trap breaks the translational invariance. Inside a trap, the atomic densities vary in space, which, in a local density approximation, translates to a local Fermi wavenumber. A spatially varying wavenumber has a direct impact on the phases, such as by smoothing the discontinuous transition between the (possible) Peierls and collapsed/separated phases. Naturally, the presence of a trap entails a larger number of involved wavenumbers and thereby a broadening of the Bragg peaks in a time-of-flight detection. Additionally, we find that the interplay between the trap, the dynamical boson field, and the fermions results in an insulator bearing a clear fermionic ``wedding cake structure'', which differs from the density profile obtained for a static periodic potential.
 
Our narrative is structured as follows. Section~\ref{sec:background} introduces the Bose-Fermi system under study and the model we employ for its description. There we also discuss some basic concepts and known results pertaining to the properties of the involved phases. The numerical methods are developed in Sec.~\ref{sec:meth}, while the associated results are presented in Sec.~\ref{sec:result}, first for the translationally invariant case~\ref{sec:trans}, and subsequently in the presence of a harmonic trap ~\ref{sec:trap}, where we focus on the quantum coherence registered in the one-particle density matrix and on the decay of spatial correlations. Concluding remarks in Sec.~\ref{sec:con} close our paper out, while the two appendixes present two possible paths to the derivation of effective model Hamiltonians taking quantum fluctuations into account for both species, alongside some details underlying the employed hybrid mean-field method.

\section{Background: model Hamiltonian and phases}
\label{sec:background}

\subsection{The Hamiltonian} \label{sec:hamiltonian}

The physical system we study is a mixture of two atomic gases, one bosonic and one fermionic. Owing to tight transverse confinement (in the sense of $\omega_\perp \gg \omega_\parallel$), excited states of the transverse modes can be neglected and the low-energy physical description is quasi-one dimensional~\cite{das_bose-fermi_2003-1,miyakawa_peierls_2004-1}. To begin with, we will take the gas to be of infinite extent in the longitudinal direction, and later consider the effect of a trap. We work with dimensionless quantities, where energies are scaled by the boson recoil energy, $E_R = \hbar^2 k^2 / 2 m_b$ with $k$ the lattice wavenumber and $m_b$ the atomic boson mass, such that half the optical lattice wavelength sets the characteristic length scale of the problem. The behavior of such a system is modeled by the Hamiltonian 
\begin{equation}\label{eq:hamiltonian}
    \begin{array}{lll}
        \hat{\mathcal{H}} & =& 
        \displaystyle{\int dx \; \hat{\psi}_b^\dagger \left( - \frac{\partial^2}{\partial x^2} + V_b(x) + \frac{g_b}{2} \hat{\psi}_b^\dagger \hat{\psi}_b \right) \hat{\psi}_b}\\ \\
        & & \displaystyle{+\int dx  \; \hat{\psi}_f^\dagger \left( -\frac{1}{r} \frac{\partial^2}{\partial x^2} + V_f(x) + g_{bf} \hat{\psi}_b^\dagger \hat{\psi}_b \right) \hat{\psi}_f},
    \end{array}
\end{equation}
where $r = m_f/m_b$ is the mass ratio between the two species, $g_b$ is the effective 1D boson-boson (BB) interaction strength, and $g_{bf}$ is the effective 1D BF interaction strength. Later on, we will assume the bosons to be harmonically confined and subjected to an optical lattice, while the fermions will be harmonically confined but assumed to have negligible interaction with the optical lattice, 
\begin{align}
V_b(x) & = \tfrac{1}{2}\omega_b^2 x^2 + V_0 \sin^2(x), \label{eq:vb} \\
V_f(x) & = r \tfrac{1}{2}\omega_f^2 x^2\label{eq:vf}.
\end{align}
The last term on the RHS of Eq.~\eqref{eq:vb} is the optical-lattice potential, also denoted as $V_{OL}=V_0\sin^2(x)$, where the distance between potential minima is $a = \pi$ in dimensionless units. At first, we remove the harmonic confinement and keep only the periodic optical lattice. To ensure that the lattice interacts only with the bosonic atoms, the light frequency should be such that it is {\it not} quasi-resonant with any of the dipole transitions -- dictated by selection rules -- of the fermionic atoms. The atomic field operators $\hat\psi_{b,f}(x)$ and $\hat\psi_{b,f}^\dagger(x)$ obey the standard commutation relations for bosonic and fermionic creation and annihilation operators. Note that the Hamiltonian commutes with both particle number operators ($\alpha = b, f$),
\begin{equation}\label{eq:particlenumbers}
    \hat{N}_{\alpha} = 
        \int dx \; \hat{\psi}_{\alpha}^\dagger(x) \hat{\psi}_{\alpha}(x). 
\end{equation}
For further use, we define the average particle number of bosons/fermions as 
\begin{equation}\label{eq:particledensity}
    \bar{n}_{\alpha} = \lim_{L \to \infty}
        \frac{1}{L} \int_{-L/2}^{L/2} dx \; \langle \hat{\psi}_{\alpha}^\dagger(x)  \hat{\psi}_{\alpha}(x) \rangle,
\end{equation}
and the density $\nu_\alpha = a \bar{n}_\alpha = \pi \bar{n}_\alpha$ as the filling of bosons/fermions per lattice site. 

Hereinafter, we assume that the bosonic gas is dense enough to form a (quasi) condensed state, but not so dense that the interaction energy becomes dominant. In terms of the 1D and 3D scattering lengths $a_{s,1D}$ and $a_{s,3D}$ this implies 
\begin{equation}
    a_{s,3D} \ll \xi \ll a_{s,1D},
\end{equation}
where the healing length of the condensate is defined as $\xi = (8 \pi a_{s,3D} \rho_b)^{-1/2}$, with $\rho_b$ the condensate density. Furthermore, our approximations are consistent with limiting our attention to optical lattice strengths satisfying $V_0 \lesssim 20 E_R$. We take the fermions to be lighter than the bosons, {\it i.e.} $r = m_f / m_b < 1$, which is typically the case for the relevant experiments, {\it e.g.} for a Li$^6$-Li$^7$ mixture~\cite{ferrier-barbut_mixture_2014-1} one has $r=0.86$ and for a Li$^6$-Cs$^{133}$ mixture~\cite{desalvo2017observation,repp2013observation} one instead gets $r \approx 0.04$. We will assume $g_b > 0$ but allow any sign and magnitude of $g_{bf}$. Both the BB and BF interactions should be possible to vary in the experiment by deploying Feshbach resonances~\cite{stan2004observation,best2009role,kawaguchi2012spinor,park2012quantum,ferrier-barbut_mixture_2014-1}.

\subsection{Pairing phase and Peierls instability} \label{sec:pairing}

For low bosonic densities, the occurrence of a so-called {\it pairing phase} for BF mixtures with attractive BF interaction is well known. For weak interaction strength and light fermions, the development of modulated densities can be understood from the fact that the fermion-mediated long-range interaction between two bosons is an ultracold atomic analog to the RKKY interaction~\cite{ruderman1954indirect,mering_fermion-mediated_2010-1,de2014fermion}. These spatial modulations of the boson density alter the potential landscape for the fermions, and subsequently instigate back-action between the two species. The spatial modulations will occur at the wavenumber $2 k_F$ if we assume an RKKY interaction from a ``flat'' background of fermions with a Fermi wavevector $k_F$. In ~\cite{cazalilla_instabilities_2003-1}, a 1D BF mixture without optical lattice was investigated using bosonization and a transition was identified between a two-component gapless LL phase and a gapped pairing phase with periodic density modulations. The transition was found to ``flow'' towards the BKT fixed point with the gap opening as 
\begin{equation} \label{eq:bkt-gap}
    \Delta \sim \exp(-1/\sqrt{|g_{bf} - g_{bf}^c|})\Theta(g_{bf}-g_{bf}^c),
\end{equation}
where $\Theta(x)$ is the Heaviside step function, and $g_{bf}^c$ is the critical Bose-Fermi coupling strength. A study using discretization of the spatial coordinate and the numerical DMRG method also predicted a stable LL phase for small negative $g_{bf}$~\cite{rizzi_pairing_2008-1}. For repulsive interaction ($g_{bf} > 0$), the aforementioned works found no pairing phase. However, in Ref.~\cite{miyakawa_peierls_2004-1}, a pairing phase was indeed reported upon employing the method of a random phase approximation. 

A gap opening at $2 k_F$ and the corresponding periodic modulation in the density are reminiscent of the Peierls distortion. As first demonstrated by Peierls~\cite{peierls_quantum_1955}, a metal is unstable towards lattice distortions, {\it i.e.} a small displacement of the atoms. This phenomenon can be understood by treating the atomic displacements on the mean-field level in the well-known SSH model. Opening up a gap $\Delta_P$ in the fermion dispersion at the Fermi wavevector $k_F$, as depicted in Fig.~\ref{fig:system} (a), lowers the total energy and corresponds to a lattice modulation of wavenumber $k_P = 2 k_F$, as sketched in Fig.~\ref{fig:system} (b). There are however also some discrepancies. The gap $\Delta_P$ was derived from a mean-field treatment of the ions, while the pairing gap $\Delta$ is the gap of excitations in the full quantum system. Still, it is interesting to observe that the gap in the SSH model also has an exponential form similar to \eqref{eq:bkt-gap}. Moreover, in the original work of Peierls, the instability was demonstrated for any non-zero interaction~\cite{peierls_quantum_1955}, while the pairing effect arises as a consequence of a phase transition occurring at finite interaction strength.  

An optical lattice is, on intuitive grounds, expected to stabilize the pairing phase against a collapse, since it renders the bosons less mobile by generating a larger effective mass. On the other hand, using the same argument, the lattice may also extend a possible metallic phase to larger regions in the phase diagram. We may note that for BF mixtures where both species are subject to an optical lattice, it has been reported that the pairing phase appears for both repulsive as well as attractive BF interactions~\cite{pazy_holstein_2005-1}. Furthermore, one could in principle expect a beating between the involved length scales, the Fermi wavenumber $k_F$ and the optical lattice wavenumber $k$, that could, in principle, give rise to so-called Devil's staircase structures~\cite{chanda2022devil}. 

A last remark on the periodically modulated Peierls phase is in order. We have envisioned the boson superfluid as the agent creating a dynamical lattice for the fermions. We may consider the opposite viewpoint of a condensate living in a partially dynamic background. Within this perspective, the Peierls phase is reminiscent of a supersolid~\cite{buchler_supersolid_2003-1,titvinidze_supersolid_2008-1}, {\it i.e.}, a superfluid state that has spontaneously broken the periodicity of the Hamiltonian. Somewhat similar scenarios have been studied in atomic condensates confined within optical resonators~\cite{baumann2010dicke,leonard2017supersolid}. 

\subsection{Stability of the Bose-Fermi mixture}

For sufficiently strong BF interaction, the system is unstable towards long-wavelength fluctuations. The stability of BF mixtures without an optical lattice has already been investigated in, {\it e.g.}, Refs.~\cite{das_bose-fermi_2003-1, miyakawa_peierls_2004-1}. Applying a hydrodynamic (mean-field) description of both fermions and bosons, a linear stability analysis yields the condition
\begin{equation}\label{eq:unstable}
    \bar{n}_{f} \geq \frac{g_{bf}^2}{2 g_f g_b},
\end{equation}
where the constant $g_f = \pi^2 / r$ appears as an effective fermion interaction strength. Note that the different definition compared to Ref.~\cite{das_bose-fermi_2003-1} is due to our use of rescaled dimensionless units. The fact that a high fermion density stabilizes the mixture is particular to the 1D case. Attractive interaction leads to a collapse of both species, while a repulsive interaction induces a phase separation where the bosons and fermions either avoid each other or lump together to form BF soliton mixtures~\cite{adhikari_fermionic_2005-1,tylutki2016dark,salerno2005matter}. More precisely, for repulsive boson-fermion interaction, the bosons can form a dark soliton (density dip), filled by a bright fermionic soliton. For attractive interaction, both species form a bright soliton. This is reminiscent of the Townes solitons predicted for electromagnetic waves~\cite{chiao1964self}, and recently demonstrated in ultracold atomic Bose-Bose mixtures~\cite{chen2021observation,bakkali2021realization}. Soliton solutions are known to be unstable beyond the mean-field approach~\cite{krutitsky2010dark,rubbo2012quantum} but BF mixtures can form stable self-bound systems ({\it e.g.}, in Ref.~\cite{salasnich_self-bound_2007-1},) which in 1D are stable within the mean-field approximation as well as to higher order~\cite{rakshit_self-bound_2019}. In the course of our analysis, we will find out that an optical lattice increases the range of $g_{bf}$ values for which the mixture is stable.

\section{Methods}\label{sec:meth}

\subsection{A hybrid mean-field approximation}\label{sec:mf}

To investigate the pairing phase, we employ a ``hybrid'' approach that intertwines quantum and classical dynamics, a method which has already been followed in Refs.\cite{wang_density-functional_2012-1,karpiuk_soliton_2004-1}. We note that this path bears similarities to the mean-field approximation employed to derive the SSH model. It is also akin to the DFT method \cite{wang_density-functional_2012-1}. Due to its affinity with the SSH model, we will refer to the gapped phase in the hybrid approximation as the Peierls phase. For the ground state, we adopt the {\it ansatz} 
\begin{equation}\label{eq:hybrid-ansatz}
    | \Psi \rangle = | \psi_b \rangle \otimes | \Psi_f \rangle,
\end{equation}
where $| \Psi_f \rangle$ is a general state of $N_f$ fermions, and $| \psi_b \rangle$ is a  (generalized) coherent state of the bosons satisfying \cite{solomon_structure_1999}
\begin{equation}\label{eq:coherent}
    \hat{\psi}_b(x) | \psi_b \rangle = \psi_b(x) |\psi_b \rangle,
\end{equation}
with the complex variable $\psi_b(x)$ called the {\it condensate wavefunction} or {\it order parameter}. We impose the normalization $\int_0^L dx\, |\psi_b(x)|^2 = 1$ for a finite system and thereby factor out the total boson number $N_b$. If $m_f / m_b = r \ll 1$, it is justified to assume an adiabatic evolution, similar to the Born-Oppenheimer approximation in molecular physics~\cite{worth2004beyond}, where the lighter fermions adjust approximately instantaneously according to the heavier bosons. Under such conditions, any gauge potential~\cite{larson2020conical}, characterizing non-adiabatic corrections and arising due to back-action between the condensate and the fermions, can be neglected. The fermionic part of the Hamiltonian can then be diagonalized while keeping the bosonic degrees of freedom fixed, thus acting as an ``adiabatic potential''.  The approximation is equivalent to replacing the bosonic field operators with the condensate wavefunction in the Hamiltonian, yielding the hybrid operator
\begin{equation} \label{eq:hybrid-hamiltonian}
    \hat{\mathcal{H}}[\psi_b] = \mathcal{E}_b[\psi_b] + \sum_n \epsilon_{f,n} \hat{\psi}_{f,n}^\dagger \hat{\psi}_{f,n},
\end{equation}
where the mean-field energy functional of the bosonic part is
\begin{equation}\label{eq:bosonic-energy-functional} 
     \mathcal{E}_b[\psi_b] = N_b \int_0^L dx \; \psi_b^* \left( -\frac{\partial^2}{\partial x^2} + V_b + \frac{g_b N_b}{2} |\psi_b|^2 \right) \psi_b,
\end{equation}
and the second part is just the fermionic part of the Hamiltonian written in diagonal form, $\epsilon_{f,n}$ being the eigenvalues of the Hartree equation for the fermion orbitals, 
\begin{equation}\label{eq:hartree}
    \left( -\frac{1}{r} \frac{\partial^2}{\partial x^2} + g_{bf} N_b |\psi_b|^2 \right) \phi_{f,n} = \epsilon_{f,n} \phi_{f,n}.
\end{equation}
Note that the solution of this Hartree equation exhibits a functional dependence on the boson density, hence from now on we will be writing $\epsilon_{f,n} = \epsilon_{f,n}[\psi_b]$. From Eq.~\eqref{eq:hybrid-hamiltonian}, a nonlinear Schr\"{o}dinger equation for $\psi_b$ can be derived,
\begin{equation}\label{eq:nonlinSch}
    i \dot{\psi}_b = \left( -\frac{\partial^2}{\partial x^2} + V_b + g_{b} N_b |\psi_b|^2 + g_{bf} n_f \right) \psi_b,
\end{equation}
where the expectation value of the fermion density can be calculated as
\begin{equation}\label{eq:nf}
    n_f(x) = \langle \hat{\psi}_f^\dagger(x) \hat{\psi}_f(x) \rangle = \sum_{n=1}^{N_f} |\phi_{f,n}(x)|^2.
\end{equation}
To determine the ground state, we resort to Eq.~\eqref{eq:nonlinSch} and numerically propagate an initial condensate wavefunction in imaginary time, employing the split-operator method~\cite{feit_solution_1982}. This procedure amounts to solving the eigenvalue problem~\eqref{eq:hartree} at each time step. 

To assess the Peierls phase for a system of infinite extent, we will use the following prescription: Assume a filling $\nu_f = 1/2$, so that we expect density modulations with a period of two sites. Since the modulations are commensurate with the optical lattice, the entire problem is periodic and the solutions of the Hartree equation are Bloch waves, with a reduced unit cell in reciprocal space due to the doubled periodicity in the Peierls phase. Again, for the system evolution, we propagate in imaginary time, solving the Hartree equation at each time instant. The case of incommensurate fermion filling is interesting in its own right, as it may lead to coexisting spatially separated regions of commensurable and incommensurate phases~\cite{molina2007commensurability}. However, our method does not allow for an incommensurate filling at present, whence we leave the problem aside for later investigation.

For the Peierls phase, we appeal to the same order parameter as in the SSH model, namely the energy gap of fermionic excitations above the Fermi surface,
\begin{equation}\label{eq:peierlsgap}
    \Delta_P := \lim_{\epsilon \to 0^+} \Big( E_f(k_F + \epsilon) - E_f(k_F - \epsilon) \Big),
\end{equation}
which in turn is identical to the gap of the hybrid Hamiltonian \eqref{eq:hybrid-hamiltonian}. As we discuss in Sec.~\ref{sec:pairing}, this quantity should approximate the excitation gap of the entire system.
\begin{figure*}[!ht]
    \centering
    \includegraphics[width=0.99\textwidth]{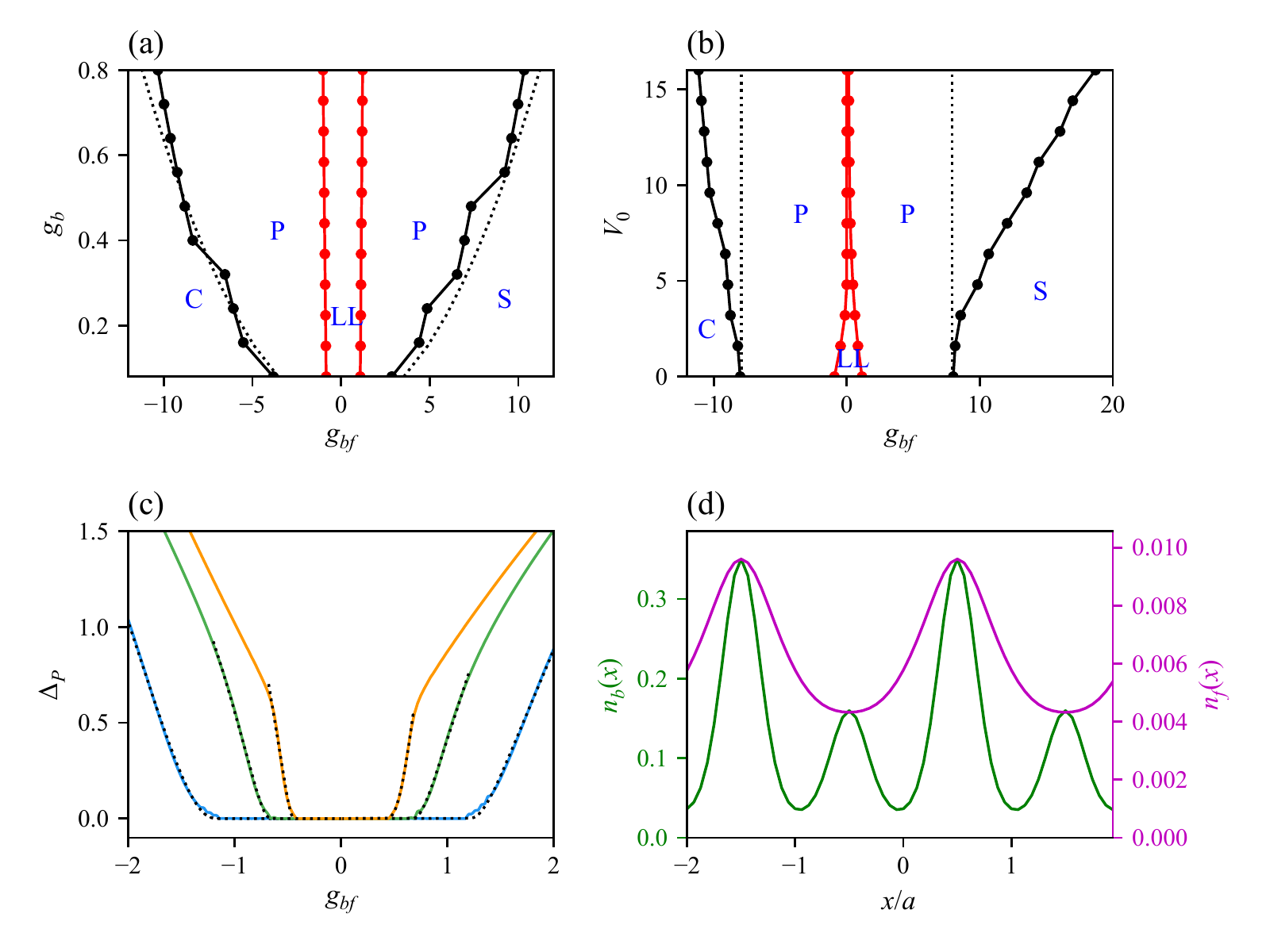}
    \caption{(color online) {\it Phase properties in the absence of a trap.} {\bf (a)} Phase diagram in the $(g_{bf}, g_b)$ plane, marking the different phases: LL, Peierls, collapsed, and separated. The solid black lines mark the boundaries of first-order transitions, while the solid red lines represent the BKT transitions. The mean-field prediction~\eqref{eq:unstable} is plotted with dotted black lines. The critical coupling $g_{bf}^c$ for the LL-to-Peierls transition is calculated from a least square fit of the Peierls gap to the expression~\eqref{eq:bkt-gap}, using the hybrid mean-field method. {\bf (b)} Same as for (a) but in the $(g_{bf}, V_0)$ plane. Here it becomes clear that the presence of the optical lattice alters the phase diagram quantitatively, in particular through stabilizing the Peierls phase. {\bf (c)} The Peierls gap as a function of $g_{bf}$, once more calculated using the hybrid code, for $V_0 = 0$ (yellow), $V_0 = 4$ (green), and $V_0 = 16$ (blue), fitted to the BKT formula~\eqref{eq:bkt-gap} (dotted lines). {\bf (d)} Boson and fermion densities in the Peierls phase for the parameters $V_0 = 1$, $ g_b = 0.4$, and $ g_{bf} = -1$. In all plots, we consider half filling, $\nu_f=1/2$.}
    \label{fig:fig2}
\end{figure*}
\begin{figure*}[!ht]
    \centering
    \includegraphics[width=0.99\textwidth]{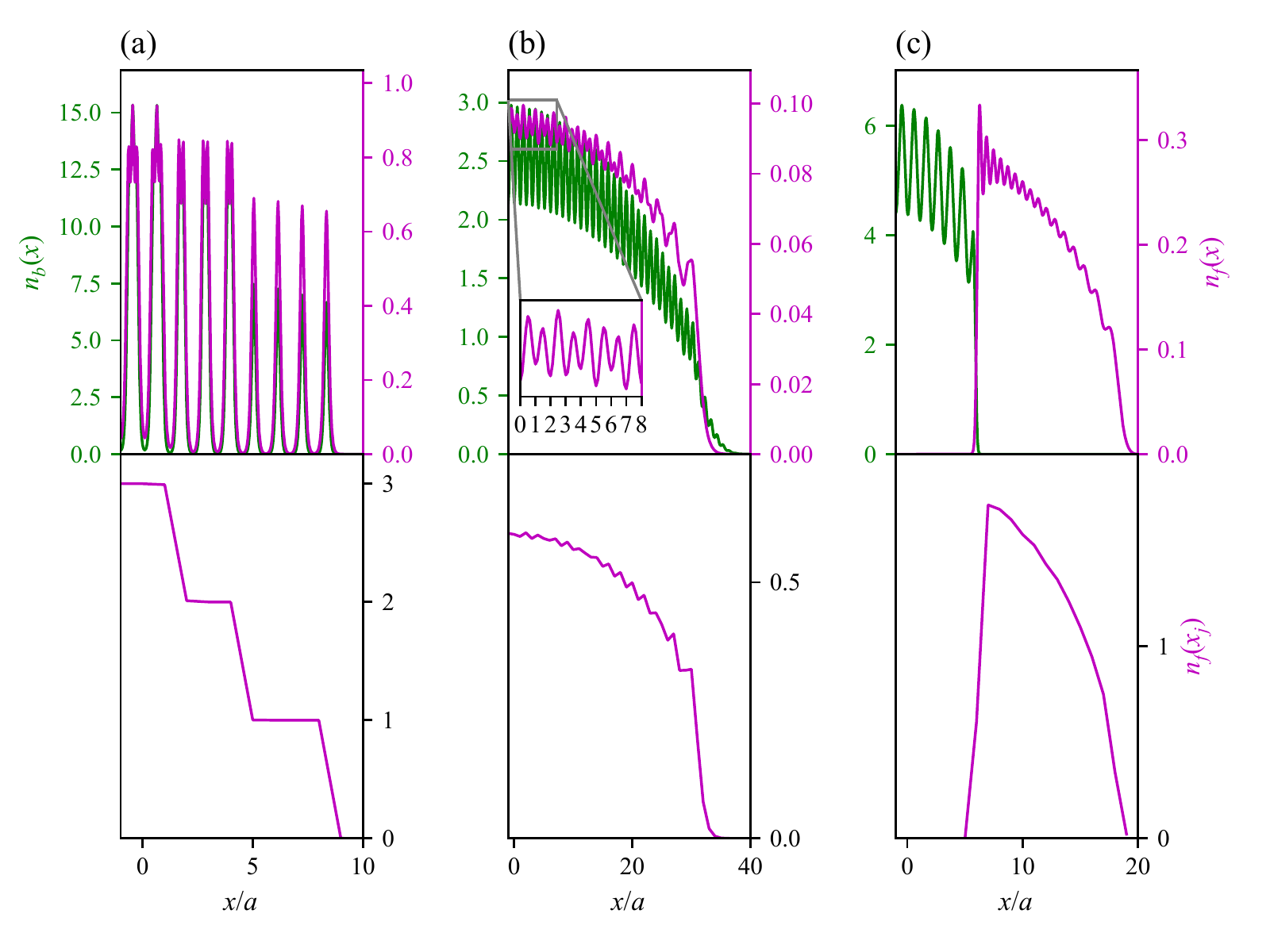}
    \caption{(Color online) {\it Density profiles in the presence of a harmonic trap.} The top row shows the boson density (magenta) and the fermion density (green), while the bottom row shows the fermion density per site. Only $x > 0$ is shown since the densities are symmetric around $x = 0$. Both plots pertain to the ground state in a BF mixture in a trap, obtained using the hybrid method. {\bf (a):} ``Wedding-cake'' structured fermionic insulator, with parameters $V_0 = 4$, $g_{bf} = -48$, $\omega_b = 0.03$ and $\omega_f = 0.3$. {\bf (b):} Peierls phase, with parameters $V_0 = 1$, $g_{bf} = -4$, $\omega_b = 0.072$ and $\omega_f = 0.0072$. The inset in the upper plot zooms on the density profile in the region $0 \leq x/a \leq 8$, evincing density modulations in the bulk, with their origin in the Peierls instability. {\bf (c):} Separated phase with parameters $V_0 = 4$, $g_{bf} = 4$, $\omega_b = 0.08$ and $\omega_f = 0.8$. In all simulations, we have set the remaining parameters to $g_b = 0.4$, $r = 0.04$, $N_f = 32$, and $N_b = 400$. 
    }
    \label{fig:fig3}
\end{figure*}

\subsection{Thomas-Fermi model for the fermions}\label{sec:tf}

To ascertain the transition to the collapsed or separated phases, we resort to a Thomas-Fermi model for the fermions (not to be confused with the Thomas-Fermi approximation); for an introduction, see Refs.~\cite{parr_density-functional_1989,spruch_pedagogic_1991} and for applications, see Refs.~\cite{march_non-local_2000-1,salasnich_effective_2002-1,salasnich_self-bound_2007-1}. We define a classical field $\psi_f(x) = \sqrt{n_f(x) / N_f}$, called the {\it Thomas-Fermi wavefunction}, where $n_f(x)$ has been defined in Eq.~\eqref{eq:nf} under the normalization $\int dx\, |\psi_f|^2 = 1$. The kinetic energy is approximated as a functional 
\begin{equation}
    \mathcal{T}[\psi_f] = \int dx \; \frac{\pi^2 N_f^3}{3 r} |\psi_f|^6.
\end{equation}
This form can be motivated by means of a dimensional analysis, and is derived in detail in \cite{parr_density-functional_1989,spruch_pedagogic_1991}. We obtain the Thomas-Fermi energy functional as 
\begin{equation}
    \mathcal{E}[\psi_b, \psi_f] = \mathcal{E}_b[\psi_b] + \mathcal{E}_f[\psi_b, \psi_f],
\end{equation}
where the first term on the right is given by \eqref{eq:bosonic-energy-functional} and the second by
\begin{equation}
\begin{array}{lll}
\mathcal{E}_f[\psi_b, \psi_f] & = & \displaystyle{ N_f \int dx \Bigg( \frac{\pi^2 N_f^2}{3 r} |\psi_f|^6 + \frac{1}{r} \left|\frac{\partial\psi_f}{\partial x}\right|^2}\\ \\ & & + g_{bf} N_b |\psi_b|^2 |\psi_f|^2 \Bigg).
\end{array}
\end{equation}
Taking the functional derivative, we find an equation of a two-component Gross-Pitaevskii type, 
\begin{equation}\label{eq:two-component-gp}
\begin{array}{l}
\displaystyle{i \partial_t \psi_b = \left( -\frac{\partial^2}{\partial x^2} + V_{\text{OL}} + g_b N_b |\psi_b|^2 + g_{bf} N_f |\psi_f|^2 \right) \psi_b}, \\ \\
\displaystyle{i \partial_t \psi_f = \left( -\frac{1}{r} \frac{\partial^2}{\partial x^2} + \frac{\pi^2 N_f^2}{r} |\psi_f|^{4} + g_{bf} N_b |\psi_b|^2 \right) \psi_f.}
\end{array}
\end{equation}
Employing once more the split-operator method~\cite{feit_solution_1982}, we then evolve an initial state in imaginary time to determine the ground state. Due to the mass separation between the two species, one may argue that an adiabatic elimination of the fermion field should be justified. This would result in an effective bosonic model where the fermion-mediated RKKY-like boson-boson interactions would appear. Such an approximation, however, has a negligible benefit to the numerical solution, whence we solve the full mean-field model formulated by Eqs.~(\ref{eq:two-component-gp}).   

\section{Results}\label{sec:result}

\subsection{Translation-invariant BF mixture}\label{sec:trans}

We first focus on the periodic case in the absence of a trap, where we set $V_{\rm{trap}} = 0$. This idealization targets the thermodynamic limit in the system response, which is meaningful when assessing the universal properties of the phase transitions reported herein. A Bose-Fermi mixture with no lattice was studied in Ref.~\cite{cazalilla_instabilities_2003-1}, where it was found that, for negative $g_{bf}$, a BKT transition separates a LL from a Peierls phase; that gap had the form dictated by Eq.~\eqref{eq:bkt-gap}.

In the upper two frames of Fig.~\ref{fig:fig2}, we depict the numerically extracted phase diagrams; frame (a) shows a region in the $(g_{bf}, g_b)$ plane when keeping $V_0 = 0$ and $r = 0.04$ constant, while frame (b) displays a region in the $(g_{bf}, V_0)$ plane for $g_b = 0.4$ and $ r = 0.04$ constant. Both (a) and (b) show the region of stability of the Peierls phase towards collapse or separation (black solid lines), located via a calculation using the Thomas-Fermi approximation and compared to the mean-field prediction \eqref{eq:unstable} (black dotted lines). The transition is identified by detecting the discontinuity in the overlap between boson and fermion densities. The LL-to-Peierls transition (red solid lines) is calculated via the hybrid method, and the critical coupling $g_{bf}^c$ is obtained from a least square fit of the expression~\eqref{eq:bkt-gap} to the numerical data. From frame (a) we conclude that the region of stability has an extent which abides by the inequality \eqref{eq:unstable}, but more importantly that our method predicts a Peierls ({\it i.e.} pairing) phase for $g_{bf} > 0$, in disagreement with Ref.~\cite{cazalilla_instabilities_2003-1}. The reason for such a disparity could be simply attributed to the fact that the hybrid method entails the unphysical assumption that the solution is periodic, which hinders the detection of long-wavelength fluctuations. From the literature on Bose-Fermi-Hubbard systems, where both gases are subject to an optical lattice, we expect however an optical lattice to stabilize the Peierls phase for repulsive interaction \cite{mathey_commensurate_2007}. From frame (b) we infer that this is indeed the case: the optical lattice stabilizes the Peierls phase beyond the boundary set by condition~\eqref{eq:unstable} for a homogeneous configuration. More precisely, the extent of this phase is growing with the optical lattice potential while the LL, collapsed, and separated phases are shrinking for the shown parameter ranges. Furthermore, the Peierls gap also grows with increasing lattice amplitude, meaning that it should be easier to observe in an experiment with a stronger optical lattice, provided the full system does not enter an insulating phase due to the inclusion of an optical lattice. 

To illustrate the type of transition occurring between the LL and Peierls phases, Fig.~\ref{fig:fig2}(c) shows three examples of the Peierls gaps as a function of $g_{bf}$, for $V_0 = 0$, $ 4$, and $ 8$, keeping $g_b = 0.4$ constant. As demonstrated by the dotted curves, close to the critical value, the numerically extracted gaps fit very well the analytical expression~\eqref{eq:bkt-gap} for the gap of a BKT transition. Away from the critical point, there is a notable ``knee'' feature, which occurs when the bosons only populate every other site, and solely for $V_0 \neq 0$.

Figure~\ref{fig:fig2}(d) displays the density of bosons and fermions within the unit cell, as a typical example for a particular point picked in the phase diagram: $V_0 = 1$, $ g_b = 0.4$, and $ g_{bf} = -1$. The doubling of the periodicity is in evidence. Contrary to most earlier studies, our method is capable of capturing both amplitude and phase modulations, however, we only find modulations in amplitude. For the repulsive case, $g_{bf}>0$, the situation is similar to that shown in (d), but with the two densities being ``out-of-phase'' instead. 

In order to experimentally probe the phase diagram, one must detect the order parameter or any other quantity capable of telling the different phases apart. The gap could, in principle, be extracted via pump-probe experiments~\cite{ernst2010probing,heinze2011multiband}, while one could also imagine transport experiments to differentiate insulating and conduction phases, {\it i.e.} Peierls vs. LL. As will be evident in the next section, the trap induces a varying Fermi wave number $k_F$ which makes the Peierls gap opening not that pronounced. In a time-of-flight detection, for instance, the Bragg peaks are smeared out, and the particular one corresponding to the gap is almost invisible. Hence, in a realistic setup, including the trap, time-of-flight measurements might not turn out to be the most optimal scheme. Instead, the onsite densities, which display alternating variations, could be accessed via single-site resolution detection~\cite{bakr2009quantum}. This technique has been successfully implemented for cold fermionic gases, including even detection of higher order correlators~\cite{haller2015single,boll2016spin,scherg2021observing}.

\subsection{BF mixture in a harmonic trap}\label{sec:trap}

Current experiments rely on the application of harmonic potentials for the attainment of sufficiently long trapping times. However, the trap may have a relatively small frequency yet still confine the atoms, such that the system {\it locally} experiences a periodic potential. To study these more realistic situations, we include a harmonic trap in our analysis but still consider a very tight transverse confinement and consequently a quasi-1D configuration. 
\begin{figure*}[!ht]
    \centering
    \includegraphics[width=0.95\textwidth]{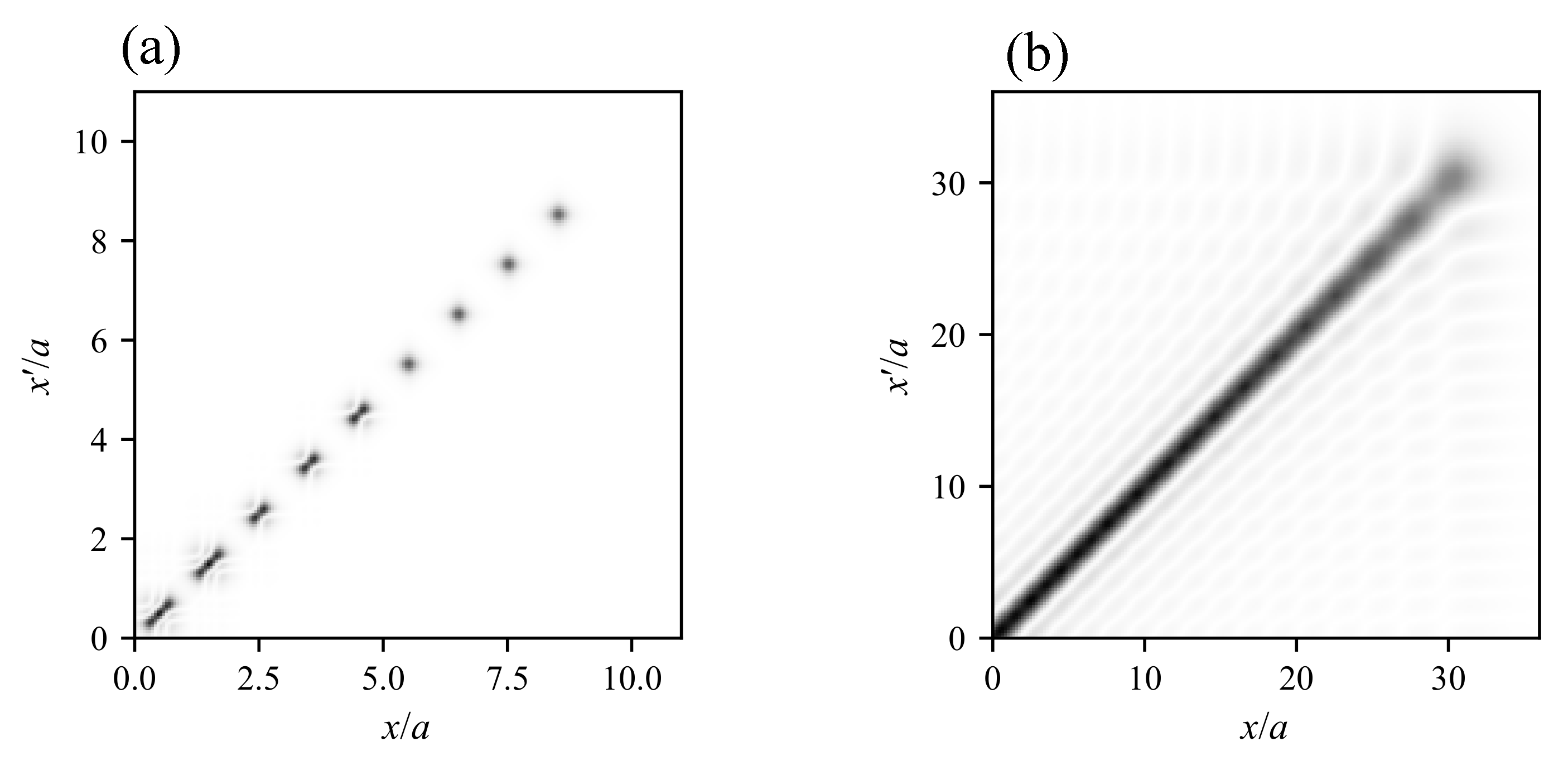}
    \caption{{\it From a Mott to a Peierls insulator with extended eigenstates in a trap.}  One-particle density matrix, $|\langle \hat{\psi}_f(x) \hat{\psi}_f^\dagger(x') \rangle|^2$, where only the top-right quadrant is plotted owing to the symmetry about $x = 0$. Two cases are shown: {\bf (a)} a trapped mixture in the {\it wedding cake} phase, for the same state as the one presented in the left column of Fig.~\ref{fig:fig3}; {\bf (b) } Peierls phase for the same state as in the central column of Fig.~\ref{fig:fig3}. The non-diagonal part of the one-particle density matrix is significantly suppressed in the wedding cake phase, where the decay is Gaussian and on a length scale similar to the characteristic length of the optical lattice. In the Peierls phase, there are instead periodic oscillations signifying coherence.}
    \label{fig:fig4}
\end{figure*}
\begin{figure*}[!ht]
    \centering
    \includegraphics[width=16cm]{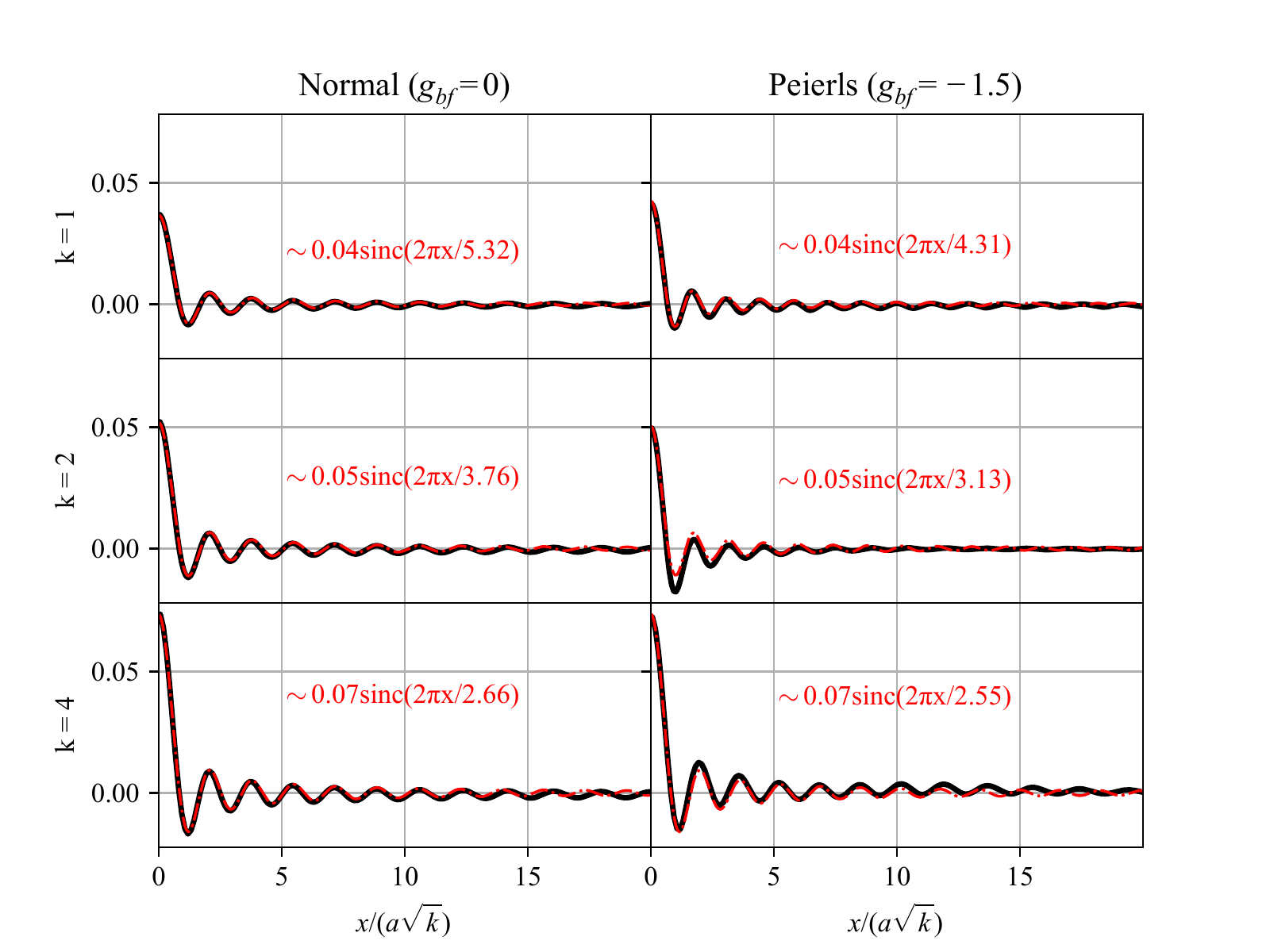}
    \caption{(Color online) {\it Scaling of the two-point density matrix with the width of the harmonic trap}. The one-point density matrix $n(x, x^\prime)$ is calculated for $x^\prime = -x$ and plotted with $x$ scaled with $a \sqrt{k}$. The oscillations of the one-point density matrix are fitted to the expression $n(x, -x) = C \text{sinc}(b x)$ in line with Eq.~\eqref{eq:asymptNf}. {\bf Left column}: normal phase ($g_{bf} = 0, \omega_{0, b} = 0.005, \omega_{0, f} = 0.07$). {\bf Right column}: Peierls phase ($g_{bf} = -1.5, \omega_{0, b} = \omega_{0, f} = 0.0036$). The harmonic confinements have been set as $\omega_{b/f} = k \omega_{0, b/f}$, with $k = 1, 2, 4$ for the top, middle and bottom rows. In all simulations, we have fixed the remaining parameters to $g_b = 0.1$ and $V_0 = 0.25$.}
    \label{fig:fig5}
\end{figure*}

We have used the hybrid method to investigate the different phases in the aforementioned setup.  Figure~\ref{fig:fig3} presents the numerical results obtained for the densities, along the main diagonal of the one-particle density matrix. In the upper row, we present the real-space densities, and in the lower row, the fermion density per site is depicted, which could, in principle, be accessed in an experiment, as mentioned in the last section. Illustrated in the left column is a novel insulating state not present in the translationally invariant system. We refer to this profile as a ``wedding cake'', drawing an analogy to the wedding cake-like structure found in the harmonically confined Bose-Hubbard model~\cite{jaksch1998cold} and also in fermionic systems~\cite{jordens_mott_2008,greif_site-resolved_2016}. In this state, the fermions form plateaus (or ``layers of the cake'') with an integer number of fermions per site. In the figure, a plateau of three fermions per site is observed in the center of the trap, transitioning in a step-like fashion to two and finally one fermion per site as we move further out from the center. The wedding cake pattern results from the interplay between kinetic, potential, and interaction energies and only appears when an optical lattice is present. It cannot be accurately modeled using a local density approximation due to the abrupt changes in density.

In the central column of Fig.~\ref{fig:fig3}, we elaborate on the correspondence to the Peierls phase occurring in the absence of a trap. Generally, the boson and fermion clouds form a region of a size determined by the trapping strength alongside the BB and BF interactions. The optical lattice induces a spatial variation in the bosonic profile, imparted to the fermions via the BF interaction. The Friedel oscillations of the fermions in this potential are then imprinted on the bosons via the RKKY interaction. Because the Peierls phase is a compressible supersolid, the wavelength varies continuously over the width of the trap. Therefore, in the trapped system, the Peierls modulations are generally not commensurate with the lattice. These modulations may be detected using single-site addressing, leading to a spatial variation in density that cannot be attributed to the profile of the trap or that of the optical lattice. We note here that the beating between the fermion and the optical lattice wavenumbers may give rise to local commensurate-incommensurate transitions in which the fermionic wavenumber of an atomic density wave spatially adjusts to the optical lattice period~\cite{molina2007commensurability}.

For any positive value of $g_{bf}$ and any strength of the optical lattice, we have observed that the system phases separate. The rightmost column of Fig.~\ref{fig:fig3} displays the separated state, where the heavier atoms occupy the center of the trap, surrounded by regions of the lighter species. Perhaps the most surprising feature of Fig.~\ref{fig:fig3} is the emergence of the wedding cake phase where we would expect a collapsed phase. The integer filling per site clearly indicates a fermionic Mott insulator state~\cite{rigol2003local,rigol2004quantum,schneider2008metallic}. One way to ascertain whether this instance is something more than a coincidence is to examine the off-diagonal elements of the one-particle density matrix, given by
\begin{equation}
    \langle \hat{\psi}_f^\dagger(x) \hat{\psi}_f(x') \rangle = \sum_{n=0}^{N_f-1} \phi_{f,n}^*(x) \phi_{f,n}(x').
\end{equation}
To derive the above expression, we have written 
\begin{equation}
    \hat{\psi}_f(x) = \sum_n \phi_{f,n}(x) \hat{\psi}_{f,n},
\end{equation}
where $\hat{\psi}_{f,n}$ are fermionic operators and the orbitals $\phi_{f,n}(x)$ are solutions of the Hartree equation (see Appendix~\ref{app:B})
\begin{equation}\label{eq:hartree-2}
    \left( -\frac{1}{r} \frac{\partial^2}{\partial x^2} + g_{bf} N_b |\psi_b|^2 \right) \phi_{f,n} = \epsilon_{f,n} \phi_{f,n}.
\end{equation}
An insulating state is expected to have an exponentially decaying one-particle density matrix~\cite{cloizeaux_analytical_1964,he_exponential_2001}. However, it turns out that this is not the case for the BF mixture in a harmonic trap.

In Fig.~\ref{fig:fig4}, we focus into a more detailed comparison between the wedding cake [frame (a)] and Peierls [frame (b)] phases, showcasing the corresponding one-body density matrix. As seen in Fig.~\ref{fig:fig3}(a), the wedding cake is distinguished by having an integer number of fermions per site. The one-particle density matrix provides an additional way of distinguishing the phases. In the wedding-cake phase, it displays a Gaussian decay on a length scale similar to the characteristic length of the optical lattice. This feature, in turn, signals the opening of a more pronounced gap in the spectrum and the presence of an insulating phase. On the other hand, the spatial decay of the matrix elements in the Peierls phase reveals a sustained coherence originating from the individual trapped-particle wavefunctions, which we will attempt to elucidate.

A characteristic example of the spatially decaying one-point density matrix $n(x, x^\prime)$ is plotted in Fig.~\ref{fig:fig5} for $x^\prime = -x$. It is evident that in the normal phase, the ``wavelength'' of the fitted sinc function scales with $\sqrt{k}$, while the relation is not so direct in the Peierls phase. To motivate the mechanism underlying the periodically decaying oscillations, we start by setting $g_{bf} = 0$, in which case the fermions are free from their interaction with the bosons in the same trap. In this case, we find (see Appendix~\ref{app:C} for further details) 
\begin{equation}\label{eq:SincCorr}
   n(x,-x)=\frac{1}{\pi} \frac{\sin(\sqrt{2N}\,2x)}{2x}.
\end{equation}
Equation~\eqref{eq:SincCorr} explicitly reveals an algebraically decaying envelope and a $\sqrt{N_f}$-dependence of the spatial frequency, with $N_f$ being the number of fermions confined in the trap, both attributes evidenced in the left plots of Fig.~\ref{fig:fig5}. While gapped phases might suggest an exponential decay of the single-particle density~\cite{sachdev1999quantum}, the spatial confinement imposed by the trap implies that the extent of excitations is typically of the same order as the characteristic trap length (unless very weak trap frequencies are considered). Thus, we observe an algebraic decay instead of an exponential one. Turning on an attractive Bose-Fermi interaction $g_{bf}$ to access the regime of Peierls instability in the right frames of Fig.~\ref{fig:fig5} marks the departure from the sinc profile and the $\sqrt{N_f}$ scaling, yet the periodicity is still evident, and the frequency carries on increasing with the extent of the trap. The counterdiagonal now defines a particular cut through ripples visible in Fig.~\ref{fig:fig4}(d) – remnants of quantum interference developing symmetrically to the main diagonal and signifying a Peierls phase with a larger spatial extent than the normal. The overall amplitude of the density matrix along the counterdiagonal, depicted in the right frames of the figure, also increases with $N_f$, in line with the prediction of Eq.~\eqref{eq:asymptNf}. Taking $N_f \to \infty$, we obtain the delta function as the familiar limit of the sinc function, writing $n(x,-x) \propto \delta(x)$.

\section{Concluding remarks}\label{sec:con}

In this work, we have studied the response of a dilute mixture of two atomic clouds of disparate masses, one fermionic and one bosonic, to an external lattice potential alongside an atomic trap. The gases are confined in a cylindrical trap so that the low-energy physical behavior of the system is effectively one-dimensional, while the optical lattice imprints a period structure onto the bosonic gas. We have assumed a weak optical lattice and a high density of bosons, such that the bosons form a condensate, which amounts to an effective interaction of the fermions with a coherent field. 

We confirmed that a Peierls instability, manifested via a BKT-type phase transition to a Peierls phase, persists when the optical lattice potential is applied solely to the bosons. Moreover, the presence of this phase was found to stabilize the system against collapse and separation, as well as to enhance the significance of the Peierls effect, a property that is highly desirable for experimental explorations of the associated phonon-like physics. The presence of the Peierls phase was confirmed upon developing and applying a hybrid mean-field method, which retained an amount of coherence able to sustain sinc-type oscillations along the counterdiagonal of the one-particle density matrix. It is interesting to note that such a method can reveal the BKT nature of the phase transition, as well as the similarity to the SSH model of the Peierls instability. The crucial scaling is captured by the method we have developed owing to the fact that we take the deformation of the Fermi surface into account, while we also retain the quantum degrees of freedom for the fermions albeit within an adiabatic approximation. An algebraic decay of correlations is found for all phases occurring in the trapped BF mixture, in accordance with a BKT phase transition.  

The natural step forwards would be to take quantum fluctuations of the boson degrees of freedom into account. This entails the development of a formulation and related numerical methods manifestly beyond any mean-field reduction, and to that aim we outline possible directions for deriving the corresponding many-body Hamiltonian in Appendix~\ref{sec:quant} below. Another promising direction for future research concerns the investigation of fermion fillings incommensurate with the optical lattice as well as studying the quenched time-evolution problem of a single fermion in a dynamical bosonic potential. Furthermore, an interesting extension replaces the optical lattice with a BB mixture, using either trapped ions or dipole bosons. Finally, systems of the kind can be analyzed in greater detail using matrix product states or bosonization techniques.

\appendix

\section{Routes towards known quantum many-body models}\label{sec:quant}

In Sec.~\ref{sec:mf}, we reported on the presence of a Peierls phase at a hybrid mean-field level. We argue that this description accounts for the pairing effect qualitatively. However, there are clear limitations to such an approach. Effectively, the model is a free fermion theory, and as long as there is a finite Peierls gap, $\Delta\neq0$, however we find a nontrivial insulator with algebraically decaying order for the trapped BF mixture. This is in line with what one expects from a BKT transition, where one phase should exhibit algebraic decay. Furthermore, the classical field stemming from the boson condensation cannot build up quantum correlations with the fermions, while a quantum field can. To quantitatively analyze such situations one must go beyond mean-field. In this appendix we outline, without going into details, two possible approaches in order to derive effective Hamiltonians for which both species are treated quantum mechanically.

In Ref.~\cite{bruderer_polaron_2007-1}, the condensate wavefunction was calculated for zero coupling between the two species, and subsequently, Bogoliubov excitations around the mean-field response were taken into account, arising from a weak interaction. Due to the presence of a deep external lattice potential in their model, a Hubbard-Holstein model is obtained which can be understood in terms of polarons. In our case, we instead find the fermion-phonon Hamiltonian
\begin{align}
     \hat{\mathcal{H}} = & \hat{\mathcal{H}}_f + \sum_{\mu} \omega_\mu \int dx \; \left( M_{\mu} \hat{b}_\mu + M_{\mu}^* \hat{b}_\mu^\dagger \right) \hat{\rho}_f \nonumber \\
     & + g_{bf} \int dx |\phi_b|^2 \hat{\rho}_f + \sum_\mu \omega_\mu \hat{b}_\mu^\dagger \hat{b}_\mu,
\end{align}
where
\begin{align}
    M_\mu(x) = u_\mu(x) - v_\mu(x)
\end{align}
are given in terms of the $u_\mu(x), v_\mu(x)$ of the Bogoliubov transformation and $\omega_\mu$ are the energies of the Bogoliubov modes. The third term will confine the fermions to their lowest Bloch band if the interaction is strong enough, and in this limit, the model approaches the Hubbard-Holstein model of Ref.~\cite{bruderer_polaron_2007-1}. However, the approach above would have to be modified since it rests on the assumption that $g_{bf}$ is weak. 

For a deep optical lattice one could come up with another method: expand the boson field in the corresponding Wannier functions, $w_j(x)$ localized at site $j$, as~\cite{lewenstein2007ultracold,Bloch2008}
\begin{equation}
    \hat\psi_b(x)=\sum_j\hat a_jw_j(x).
\end{equation}
If we expand the fermion field as $\hat\psi_f(x)=\sum_k\hat c_ke^{ikx}$ and impose the single-band and tight-binding approximations, we arrive at the many-body Hamiltonian
\begin{align}
    \hat{\mathcal{H}}_{mb} & = \hat{\mathcal{H}}_{BH} +\sum_k\!\frac{k^2}{r}\hat c_k^\dagger\hat c_k \nonumber \\
    & +\hat N_b\sum_{k,k'}\!\left[D(k-k')\hat c_k^\dagger\hat c_{k'}+\text{H.c.}\right]\!,
\end{align}
where
\begin{align}
\hat{\mathcal{H}}_{BH} & =-J\sum_j\left(\hat a_j^\dagger\hat a_{j+1}+\text{H.c.}\right) \nonumber \\
& +\frac{U}{2}\sum_j\hat n_j\left(\hat n_j-1\right)
\end{align}
is the Bose-Hubbard Hamiltonian, with $J$ and $U$ the tunneling rate and onsite interaction strengths, respectively, $\hat n_j=\hat a_j^\dagger\hat a_j$, and $D(k-k')$ is the overlap integral (which is Gaussian in the harmonic approximation). Since $\hat N_b$ is preserved, the quadratic fermionic Hamiltonian can be readily diagonalized, and we find no back-action on the bosons due to the fermions. To incorporate such effects one would need to go beyond the single-band or tight-binding approximations. Alternatively, the approximations may be kept but other additional degrees of freedom should be introduced corresponding to the phonon modes. For heavy bosons, one can follow the idea of Ref.~\cite{maluckov2013high} to allow a variation in the position of the Wannier-function centers $j\pi$, but keep the shape of the functions intact. Thus, we associate a quantized shift $\hat{\delta}_j$ with each Wannier function, $w_j(x)\rightarrow w_j(x-\hat{\delta}_j)$. Assuming small shifts $\hat{\delta x}_j\ll1$ one may then expand to linear order in them and derive an effective Fr\"ohlich-like Hamiltonian~\cite{mahan2013many}
\begin{equation}
    \hat{\mathcal{H}}_{Fr}=\hat{\mathcal{H}}_{mb}+\hat{\mathcal{H}}_{BP}+\hat{\mathcal{H}}_{FBP},
\end{equation}
where the boson-phonon interaction is given by
\begin{equation}
\begin{array}{lll}
    \hat{\mathcal{H}}_{BP} & = & \displaystyle{-J_1\sum_j\left(\hat\delta_{j+1}-\hat\delta_j\right)\left(\hat a_j^\dagger\hat a_{j+1}+\text{H.c.}\right)}\\ \\
    & & \displaystyle{+\frac{U_1}{2}\sum_j\hat\delta_j\hat n_j\left(\hat n_j-1\right)}.
    \end{array}
\end{equation}
If we introduce the local phonon annihilation/creation operators $\hat d_j/\hat d_j^\dagger$, the phonon displacement is expressed as 
\begin{equation}
    \hat\delta_j=\left(\hat d_j+\hat d_j^\dagger\right)/2,
\end{equation}
while the fermion-boson-phonon interaction term takes the form
\begin{equation}
    \hat{\mathcal{H}}_{FBP}=\mu\hat N_d+\sum_j\sum_{k,k'}\hat n_jD_1(k-k')\hat\delta_j\hat c_k^\dagger\hat c_{k'},
\end{equation}
where the first term is the bare phonon energy  ($\mu$ is the characteristic frequency and $\hat N_d=\sum_j\hat d_h^\dagger\hat d_j$). In principle, within the harmonic approximation, the coefficients $J_1$, $U_1$ and $D_1$ can be analytically determined. Both $\hat{\mathcal{H}}_{BP}$ and $\hat{\mathcal{H}}_{FBP}$ describe phonon-assisted tunneling, either between neighboring lattice sites (bosons) or between different momentum modes (fermions). It may be noted that if $g_{bf}=0$, then $D(k-k')=D_1(k-k')=0$ and $\hat N_d=0$ such that $\hat{\mathcal{H}}_{BP}=0$. 

\section{The hybrid mean-field--quantum method}
\label{app:B}

In this appendix, we provide some further details about the employed hybrid mean-field {\it ansatz}. As already mentioned in Sec.~\ref{sec:mf}, a gap of the form \eqref{eq:bkt-gap} opens up in the pairing phase. A superfluid of bosons is gapless and, to the lowest order, is described by the condensate wavefunction $\psi_b(x)$ -- the mean-field order parameter. We define the generalized coherent state
\begin{equation}
    \ket{\psi_b} = \exp\left( \int dx \; \psi_b(x) \hat{\psi_b}^\dagger(x) - \text{H.c.} \right) \ket{0},
\end{equation}
where $\ket{0}$ is the bosonic vacuum. Taking the expectation value of the time-dependent Schr\"{o}dinger equation, we obtain an effective hybrid Hamiltonian
\begin{align}
    & \hat{\mathcal{H}}_{\text{eff}} = \mathcal{E}_b[\psi_b] + \hat{\mathcal{H}}_f[\psi_b] - i \matrixel{\psi_b}{\partial_t}{\psi_b},
\end{align}
where the first term is given by \eqref{eq:bosonic-energy-functional}, the second term is
\begin{align}
    \hat{\mathcal{H}}_f[\psi_b] = \int dx \; \hat{\psi}_f^\dagger \Big( -\frac{1}{r} \frac{\partial^2}{\partial x^2} + V_f + g_{bf} |\psi_b|^2 \Big) \hat{\psi}_f,
\end{align}
and the third term is the Berry connection~\cite{larson2020conical}. We will assume that the state of the condensate is slowly modified, so that this term can be neglected. 

The above is equivalent to a product-state {\it ansatz} for the ground state
\begin{equation}
    \ket{\Psi_0} = \ket{\psi_b} \otimes \ket{\Psi_f},
\end{equation}
where $\ket{\Psi_f}$ is a general fermion state of $N_f$ fermions. Such an {\it ansatz} neglects any entanglement built between the bosons and fermion subsystems; correlations may actually arise within the non-condensed fraction of bosons and, more importantly, different configurations of the condensate may get entangled with the fermions. In general, such mixed quantum-classical dynamics is interesting in a much wider context in condensed matter physics, high-energy physics, and quantum gravity, and has been investigated in detail in Refs.~\cite{prezhdo_mixing_1997,caro_impediments_1999,oppenheim_objective_2020}. In Ref.~\cite{prezhdo_mixing_1997}, a multi-configurational mean-field approximation based on hybrid quantum-classical theory was developed, with the central  object being the quantum-classical distribution function -- a map from the classical phase space to the set of quantum density operators, defined as
\begin{equation}
    \hat{\rho}(\mathbf{q}, \mathbf{p}) = \sum_{ij} \varrho_{ij} |\Psi_i\rangle \langle \Psi_j | \delta(\mathbf{q} - \mathbf{q}_{ij}) \delta(\mathbf{p} - \mathbf{p}_{ij}),
\end{equation}
where $\mathbf{q}, \mathbf{p}$ are the classical generalized coordinates and momenta, and $| \Psi_i \rangle$ is a basis state of the quantum subsystem. In this formalism, ``non-diagonal'' contributions correspond to the quantum subsystem generating a coupling between different trajectories of the classical subsystems. When considering only a single trajectory, the quantum-classical distribution function can be written as a single delta-function term,
\begin{equation}
    \hat{\rho}[\psi_b'] = |\Psi_f\rangle \langle \Psi_f | \delta[\psi_b - \psi_b'],
\end{equation}
where the brackets indicate a functional dependence. In other words, the system is represented by a single point in classical phase space, evolving with a quantum density matrix of a pure state. 

We find the ground state of $\hat{H}_{\text{eff}}$ self-consistently first observing that it can be readily diagonalized by selecting  
\begin{equation}
    \hat{\psi}_f(x) = \sum_n \phi_{f,n}(x) \hat{\psi}_{f,n},
\end{equation}
where the orbitals are solutions of the Hartree (single-particle) equation
\begin{equation}\label{eq:hartree-2}
    \left( -\frac{1}{r} \frac{\partial^2}{\partial x^2} + g_{bf} N_b |\psi_b|^2 \right) \phi_{f,n} = \epsilon_{f,n} \phi_{f,n}.
\end{equation}
Second, we define a non-linear Schr\"{o}dinger equation for the condensate,
\begin{equation}\label{eq:nonlinear}
\begin{array}{lll}
    i \dot{\psi}_b & = &  \displaystyle{\frac{\delta}{\delta \psi_b^*} \matrixel{\Psi_f}{\hat{H}_{\text{eff}}}{\Psi_f}}\\ \\
    & = &  
    \displaystyle{\left( -\frac{\partial^2}{\partial x^2} + V_b + g_b N_b |\psi_b|^2 + g_{bf} n_f \right) \psi_b},
    \end{array}
\end{equation}
where
\begin{equation}
    n_f(x) = \matrixel{\Psi_f}{\hat{\psi}_f^\dagger(x) \hat{\psi}_f(x)}{\Psi_f} = \sum_n |\phi_{f,n}(x)|^2
\end{equation}
depends explicitly on the bosons through \eqref{eq:hartree-2}. The ground state of \eqref{eq:nonlinear} is found from imaginary time propagation utilizing the split-operator method \cite{feit_solution_1982}. 

As mentioned in Sec.~\ref{sec:mf}, the above method can also be easily adapted to study a translationally invariant system of infinite extent for fermion filling $\nu_f = 1/2$. In this case, we assume a periodic solution $\psi_b(x + 4 \pi) = \psi_b(x)$. The solutions of the Hartree equations are then Bloch waves $e^{i q x/2} u_{f,n}(q, x)$. Due to the assumed double periodicity of $4 \pi$, the first Brillouin zone is halved and the fermionic filling is in direct correspondence to the edges of the Brillouin zone. It is straightforward to find the fermion density as the integral 
\begin{equation}
    n_f(x) = \int_{1BZ} dq \; |u_{f,0}(q, x)|^2.
\end{equation}
Another important difference here is that we have to diagonalize the Bloch/Hartree equation for each quasi-momentum -- in practice for a large number of sampled values. Apart from these discrepancies between the two methods, the solution to Eq.~\eqref{eq:nonlinear} is also found using imaginary time propagation.

\section{Decay of correlations in the absence of Bose-Fermi interaction}
\label{app:C}

For $g_{bf}=0$ (normal phase), the one-point density matrix can be calculated analytically via the Cristoffel-Darboux formula, involving summed products of individual orthogonal fermionic wavefunctions in terms of the Hermite polynomials $H_n(x)$ (with $n=0,1,\ldots N_f$) derived from the quantum harmonic oscillator eigenstates. Letting $N_f=N$ for convenience, the result reads
\begin{align}\label{eq:nexact}
    n(x, x')& = \pi^{-1/2} \exp\left(-\frac{x^2 + (x')^2}{2}\right)  \nonumber \\
    & \times\frac{1}{N! 2^{N+1}} \frac{H_N(x') H_{N+1}(x) - H_N(x) H_{N+1}(x')}{x - x'}.
\end{align}
For $x^{\prime}=-x$, along the counterdiagonal of the density matrix that we select for Fig.~\ref{fig:fig5}, the above expression can be recast into the following form:
\begin{equation}\label{eq:asymptNf}
\begin{aligned}
    &n(x, -x) = \pi^{-1/2} \exp(-x^2) \frac{1}{N! 2^{N+1}}\\
    &\times \frac{2(-1)^NH_N(x) H_{N+1}(x)}{2x} \sim \frac{2^{2N}\Gamma((N+1)/2) \Gamma(N/2+1)}{\pi^{3/2}\,\,2^N\,N!} \\ &\times\frac{\sin(\sqrt{2N}\,2x)}{2x}=\frac{1}{\pi} \frac{\sin(\sqrt{2N}\,2x)}{2x},
\end{aligned}
\end{equation}
which follows from the familiar asymptotic $N\gg 1$ expansion of quantum harmonic oscillator eigenstates~\cite{AbramowitzStegun} 
\begin{equation}
    e^{-x^2/2}H_N(x) \sim \frac{2^N}{\sqrt{\pi}} \Gamma\left(\frac{N+1}{2}\right) \cos(x\sqrt{2N}-N\pi/2)
\end{equation}
and the Legendre duplication formula for the gamma function. In fact, the number of fermions confined in the trap need not be large. Even for $N_f \gtrsim 10$ we note a reasonable agreement between the exact formula of Eq.~\eqref{eq:nexact}, and the asymptotic sinc profile of~\eqref{eq:asymptNf}. 

\begin{acknowledgments}
We thank Maciej~Lewenstein, Hannes~Conners, and Christophe~Salomon for helpful discussions and comments. We acknowledge financial support from VR-Vetenskapsr\aa det (The Swedish Research Council), and KAW (The Knut and Alice Wallenberg foundation). Th.~K.~M. acknowledges support from: ERC AdG NOQIA; Ministerio de Ciencia y Innovation Agencia Estatal de Investigaciones (PGC2018-097027-B-I00/10.13039/501100011033, EX2019-000910-S/10.13039/501100011033, QUANTERA DYNAMITE PCI2022-132919, Proyectos de I+D+I ``Retos Colaboración'' QUSPIN RTC2019-007196-7); MICIIN with funding from European Union NextGenerationEU(PRTR-C17.I1) and by Generalitat de Catalunya; Fundació Cellex; Fundació Mir-Puig; Generalitat de Catalunya (European Social Fund FEDER and CERCA program, AGAUR Grant No. 2021 SGR 01452, QuantumCAT \ U16-011424, co-funded by ERDF Operational Program of Catalonia 2014-2020); Barcelona Supercomputing Center MareNostrum (FI-2022-1-0042); EU Horizon 2020 FET-OPEN OPTOlogic (Grant No 899794); EU Horizon Europe Program (Grant Agreement 101080086 — NeQST); ICFO Internal ``QuantumGaudi'' project; European Union's Horizon 2020 research and innovation program under the Marie-Skłodowska-Curie grant agreement No 101029393 (STREDCH) and No 847648 (``La Caixa'' Junior Leaders fellowships ID100010434: LCF/BQ/PI19/11690013, LCF/BQ/PI20/11760031, LCF/BQ/PR20/11770012, LCF/BQ/PR21/11840013). Views and opinions expressed in this work are, however, those of the author(s) only and do not necessarily reflect those of the European Union, European Climate, Infrastructure and Environment Executive Agency (CINEA), nor any other granting authority. Neither the European Union nor any granting authority can be held responsible for them.
\end{acknowledgments}

\bibliography{references}{}
\bibliographystyle{unsrtnat}

\end{document}